\documentclass{ceab}   

\usepackage{epsfig}     
\usepackage{graphicx}   

\usepackage{ceabbib}     
\usepackage[T1]{fontenc}

\def\BD{BD\,+37$^\circ$\,442}   
\def\HD{HD\,49798}
\def\ROSAT{\textit{ROSAT}}
\def\XMM{\textit{XMM--Newton}}
\def\aut{\textit{La Palombara et al.}}
\def\tit{\textit{X--ray observations of \HD\ and \BD}}

\begin{document}

\title{Observations of the unique X--ray emitting subdwarf stars \HD\ and \BD}

\author{N. La Palombara$^{1}$, S. Mereghetti$^1$, A. Tiengo$^2$, P. Esposito$^1$, L. Stella$^3$, G.L. Israel$^3$
\vspace{2mm}\\
\it $^1$INAF - IASF Milano, via Bassini 15, I--20133 Milano, Italy\\ 
\it $^2$IUSS, piazza della Vittoria 15, I--27100 Pavia, Italy\\
\it $^3$INAF - OA Roma, via Frascati 33, I--00040 Monteporzio Catone, Italy
}
\maketitle

\begin{abstract}
We report on the results we obtained with \XMM\ observations of \HD\ and \BD, the only two sdO stars for which X--ray emission has been observed so far. \HD\ is a single-lined spectroscopic binary with orbital period of 1.5 days. We could establish that its companion is a massive white dwarf with $M$ = 1.28 $M_{\odot}$, which makes it a candidate type Ia supernova progenitor; we also detected a significant X--ray emission during the white--dwarf eclipse, which could be X--ray emission of the sdO star itself. In the case of \BD, a luminous He--rich sdO that up to now was believed to be a single star, we discovered soft X--ray emission with a periodicity of 19.2 s. This indicates that also this hot subdwarf has a compact binary companion, either a white dwarf or a neutron star, most likely powered by accretion from the wind of the sdO star.
\end{abstract}


\section{Introduction}

Hot subdwarfs are evolved, low--mass stars that have lost most of their
hydrogen envelope and are now in the stage of helium--core burning
\citep{heb09}. From the spectroscopic point of view we
distinguish the cooler B--type subdwarf (sdB) stars, with effective temperature 
$T_{\rm eff} <$ 40,000 K, and the hotter O--type (sdO) stars, with $T_{\rm eff} >$ 40,000 K \citep{hir08}. Many hot subdwarfs are in close binary systems \citep{max01,nap04long,cop11}, and many of the subdwarf companions should be white dwarfs (WD) \citep{han02}. The presence of a compact companion (a WD, neutron star (NS), or black hole) could be revealed by the detection of X--ray emission powered by accretion, if the subdwarf mass donor provides a sufficiently high accretion rate. Although the fraction of binaries among sdO stars is lower than for the sdBs, the prospects to find X--ray emitting
companions are more promising for these stars, since at least a few
luminous sdO stars show evidence for stellar winds with mass loss rate $\dot M \sim
10^{-7}-10^{-10} M_{\odot}$ yr$^{-1}$ \citep{jef10}. Here we report on the results we obtained with \XMM\ observations of \HD\ and \BD\ (Table~\ref{parameters}), the only two sdO stars for which X--ray emission has been observed up to now.

\begin{table}[htbp]
\caption{Main parameters of the sdO stars \HD\ and \BD.}\label{parameters}
\vspace{-0.5cm}
\begin{center}
\begin{tabular}{lcccc} \hline \hline
Parameter				& \multicolumn{2}{c}{\HD}		& \multicolumn{2}{c}{\BD}		\\
					& Value			& Reference	& Value			& Reference	\\ \hline
$M$ ($M_{\odot}$)			& 1.50 $\pm$ 0.05	& 1		& 0.9			& 5		\\
$R$ ($R_{\odot}$)			& 1.45 $\pm$ 0.25	& 2		& 1.6			& 5		\\
log $g$					& 4.35 $\pm$ 0.20	& 2		& 4.00 $\pm$ 0.25	& 6		\\
log ($L$/$L_{\odot}$)			& 3.95 $\pm$ 0.25	& 2		& 4.5 $\pm$ 0.3		& 6		\\
$T_{\rm eff}$ (K)			& 47,500 $\pm$ 2,000	& 2		& 48,000 $\pm$ 2,000	& 7		\\
$d$ (kpc)				& 0.65 $\pm$ 0.1	& 2		& 2$^{+0.9}_{-0.6}$	& 6		\\
$v_{\infty}$ (km s$^{-1}$)		& 1,350			& 3		& 2,000			& 7		\\
$\dot M$ ($M_{\odot}$ yr$^{-1}$)	& 10$^{-8.5}$		& 4		& 10$^{-8.5}$		& 7		\\ \hline
\end{tabular}
\end{center}
\begin{small}
References: 1 - \citealt{mer09long}; 2 - \citealt{kud78}; 3 - \citealt{ham81}; 4 - \citealt{ham10}; 5 - \citealt{hus87}; 6 - \citealt{Bauer&Husfeld95}; 7 - \citealt{jef10}
\end{small}
\end{table}

\section{\HD}

\subsection{Mass measurement of the white--dwarf companion}

\begin{figure}[h!]
\centering
\resizebox{\hsize}{!}{\includegraphics[angle=0,height=1cm]{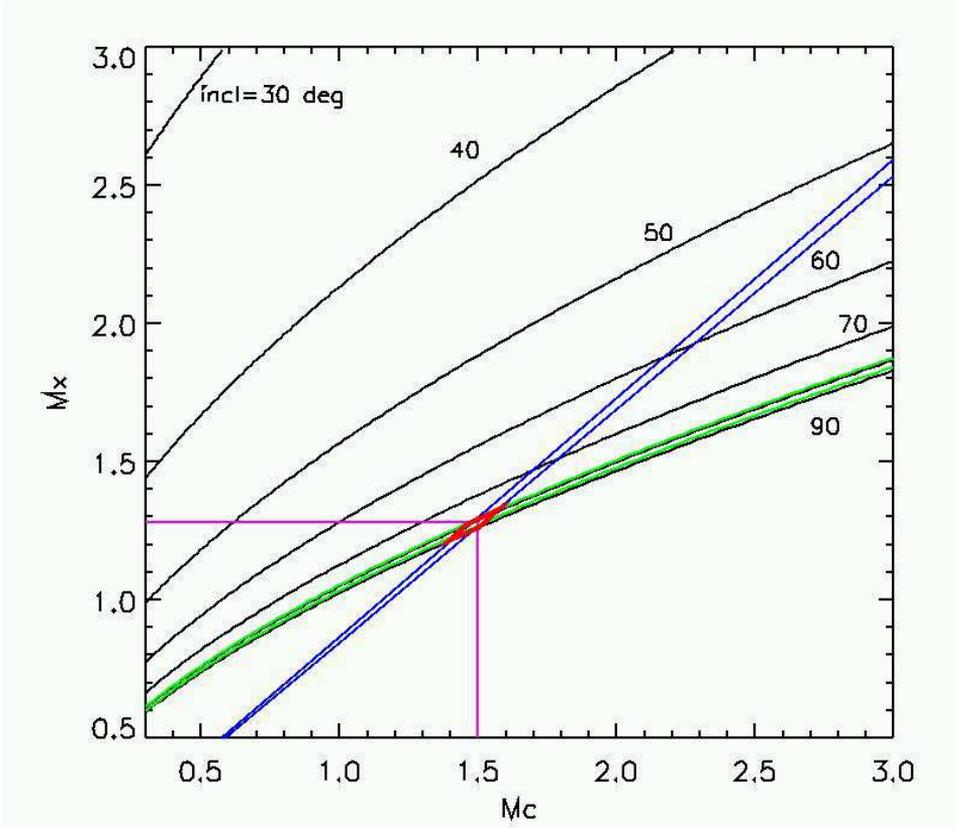}}
\caption{Estimated range of masses for the \HD\ subdwarf (M$_{\rm C}$) and its WD companion (M$_{\rm X}$) as a function of the system inclination. The two straight lines correspond to the mass ratio $q$ = 1.17 $\pm$ 0.01. The axis are in units of solar masses.}
\label{masse}
\vspace{-0.5cm}
\end{figure}

\HD\ is a single-lined spectroscopic binary, with orbital period $P_{orb}$ = 1.55 d and projected semi-major axis $A_{\rm C}$ sin($i$)$ = (2.51 \pm 0.01) \times 10^{11}$ cm \citep{tha70,sti94}. These values imply an optical Mass Function $f_{\rm opt} = 4\pi^{2}$ [$A_{\rm C}$sin($i$)]$^{3}$/[$GP_{\rm orb}^{2}$] = 0.263 $\pm$ 0.003 $M_{\odot}$ = [$M_{\rm X}$sin($i$)]$^{3}$/[$M_{\rm X}+M_{\rm C}$]$^{2}$, which provides the relation between the masses of the two stars; they are shown, for different values of inclination $i$, by the curved lines of Fig.~\ref{masse}.

In 1997 the \ROSAT\ satellite performed the first X--ray observation of this system, and discovered a soft and pulsed X--ray emission with P = 13.2 s, a high pulsed fraction ($\sim$ 70 \%) and a sinusoidal profile \citep{isr97long}. Since \HD\ is characterized by a significant stellar wind (Table~\ref{parameters}), the observed X--ray emission is most probably due to mass accretion onto a compact companion, either a WD or a NS, in the wind of the sdO star, which is modulated by the rotation of the compact object.

In 2008 a 44 ksec \XMM\ observation provided a high--statistics spectrum, which was described with the sum of a soft blackbody (BB), with temperature $kT_{\rm BB}$ = 39 $\pm$ 1 eV and radius $R_{\rm BB}$ = 16 $\pm$ 2 km, and a hard power--law (PL) model, with photon index $\Gamma$ = 1.96 $\pm$ 0.03; the corresponding source luminosity was $L_{X} \sim 10^{32}$ erg s$^{-1}$. This results confirmed that the compact accreting companion is a WD \citep{mer09long}.

The observation covered almost one third of the orbital period, therefore it was possible to measure with high precision the time delays of the X--ray pulses that were induced by the orbital motion, to fit them with a sinusoidal function and, then, to obtain the X--ray projected semi--major axis: $A_{\rm X}$ sin($i$)$ = (2.93 \pm 0.02) \times 10^{11}$ cm. Since we already knew the optical projected semi--major axis, it was also possible to estimate for the first time the mass ratio between the sdO and the WD stars: $q$ = [$A_{\rm X}$ sin($i$)]/[$A_{\rm C}$ sin($i$)] = $M_{\rm HD}$/$M_{\rm WD}$ = 1.17 $\pm$ 0.01. In this way it was possible to put a strong constraint on the system inclination and the range of allowed masses for the two stars (Fig.~\ref{masse}).

The deciding advancement in this search was provided by the observation of the eclipse of the X--ray source, since its duration depends on both the subdwarf radius and the orbit inclination: $(R_{\rm HD}/a)^{2}$ = cos$^{2}(i)$ + sin$^{2}(i)$sin$^{2}(\theta)$, where $a$ is the orbital separation and $\theta$ is the eclipse half angle. We already knew $R_{\rm HD}$, therefore the measurement of the eclipse duration ($\sim$ 1.3 hours) enabled us to get the system inclination angle: $i = 79^{\circ}-84^{\circ}$ (Fig.~\ref{masse}). In this way we obtained a definitive measurement of both the mass of the sdO star ($M_{\rm HD}$ = 1.50 $\pm$ 0.05 $M_{\odot}$) and that of its WD companion ($M_{\rm WD}$ = 1.28 $\pm$ 0.05 $M_{\odot}$). Moreover, independently of the dimensions of the subdwarf star, a lower limit of 1.2 $M_{\odot}$ was obtained for the extreme case of an inclination angle $i = 90^{\circ}$.

Compared to the distribution of the WD masses, this result makes the WD companion of \HD\ one of the most massive WDs currently known. Most WD masses are derived with indirect methods, such as surface gravity estimates (obtained by spectral modeling) or the measurement of the gravitational redshift. They depend on models and assumptions, and provide combined information on mass and radius, while our mass determination is based on a direct dynamical measurement which is independent of any modelling. Another unique property of this source is that it is the fastest spinning WD, and its short period implies a low magnetic field (of a few kG at most) in order to allow the mass accretion at the low flux level which is observed. Finally, its location in a post common--envelope binary, with well determined masses, provides an optimal test bench for the evolutionary models. The future evolution of \HD\ will most likely imply an increased mass--transfer rate onto the WD through Roche--lobe overflow, which could lead to an accreted mass of a few tenths of solar masses: in this way the WD mass would be pushed above the Chandrasekhar limit, likely triggering a type Ia supernova explosion or an accretion--induced collapse.




\subsection{Observation of the X--ray source eclipse}

Very surprisingly, also during its eclipse phase the source X--ray flux did not reduce to zero, since the source was still visible with a flux $f^{\rm ecl}_{\rm 0.2-10~keV} \simeq 4\times 10^{-14}$ erg cm$^{-2}$ s$^{-1}$, corresponding to $\simeq$ 10 \% of that observed out of the eclipse. The spectrum can be described with a thermal Bremsstrahlung model with temperature $kT \simeq 0.51$ keV, and the observed flux implies log($f_{\rm 0.5-10~keV}^{\rm unabs}/f_{\rm bol}$) $\simeq$ -7.4 \citep{mer11blong}. We note that soft X--rays are commonly observed in normal early--type stars, where shocks in the radiation--driven stellar winds are the most probable origin. In their case, a canonical linear proportion links the X--ray and bolometric fluxes: log($f_{\rm 0.5-10~keV}^{\rm unabs}/f_{\rm bol}$) = -6.45 $\pm$ 0.51 \citep{naz09}. This value is comparable to that measured during the eclipse phase of \HD: therefore it is possible that we were observing, for the first time, the intrinsic X--ray emission from a hot subdwarf, even if we cannot exclude that this flux is due to the scattering of the primary emission in the wind of \HD.

\section{\BD}

In order to check the possibility of intrinsic X--ray emission, we looked for other sdO stars similar to \HD\ but without any companion. We found that the best candidate was \BD, since, from several points of view (mass and radius, surface gravity and temperature, luminosity, wind parameters) it is similar to \HD\ (Tab.~\ref{parameters}). \BD\ is roughly 3 times more distant than \HD, but the main difference is that for this star there is no evidence of a companion, neither from optical spectroscopic \citep{fay73,kau80,dwo82} or photometric data \citep{Landolt68,Landolt73}, nor from infrared data \citep{Thejll+95}.

We obtained an observation of about 30 ks with \XMM, which provided a clear source detection \citep{LaPalombara+12}. Its spectrum is very soft and, as in the case of \HD, can be fit with a PL+BB model; while the PL is poorly constrained, the BB parameters ($kT_{\rm BB}$ = 45 $\pm$ 10 eV, $R_{\rm BB} = 39_{-28}^{+162}$ km) are comparable to those obtained for \HD. The total BB luminosity is strongly dependent on the PL parameters, but it is in the range from $6 \times 10^{31}$ to $1.5 \times 10^{35}$ erg s$^{-1}$. The source flux corresponding to the best--fit spectrum is $f^{\rm abs}_{\rm 0.2-1~keV}$ = (2.6 $\pm$ 0.3)$\times 10^{-14}$ erg cm$^{-2}$ s$^{-1}$, 28 \% of which can be ascribed to the power--law component. It implies a X--ray/bolometric flux ratio log($f_{\rm 0.5-10~keV}^{\rm unabs}/f_{\rm bol}$) = -6.63, a value very close to the canonical one for the normal O--type stars.

However, it is not possible to attribute the observed X--ray flux to the intrinsic emission in the sdO star, since the timing analysis showed that the observed X--ray emission is characterized by a periodic modulation, with a period $P$ = 19.156 $\pm$ 0.001 s. The pulse profile shows a single broad peak with a pulsed fraction of 31 $\pm$ 4 \%. No significant differences are seen in the light curve shape and pulsed fraction in different energy ranges. These timing properties cannot be explained by the shocks in the stellar winds, as in the case of the normal early--type stars, therefore this hypothesis must be rejected. The short period, the sinusoidal profile and the large pulsed fraction imply that the observed X--ray emission comes from a compact companion of the subdwarf star rather than in \BD\ itself. The periodic modulation can be explained equally well as the spin period of a NS or a WD. 

\begin{figure}[h]
\centering
\resizebox{\hsize}{!}{\includegraphics[angle=0,clip=true]{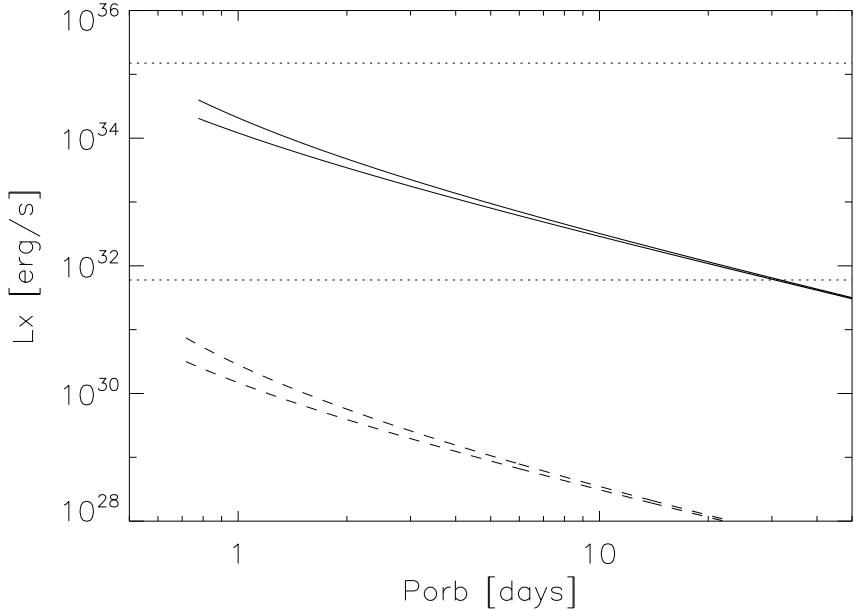}}
\caption{Estimated wind--accretion X--ray luminosity of \BD\ as a function of the binary period,
in the case of a WD (\textit{dashed lines}) and of a NS (\textit{solid lines}). The curves refer to a NS of 1.4 $M_{\odot}$ and 10 km radius, and to a WD of 0.6 $M_{\odot}$ and 10,000 km radius; both cases are referred to a mass of 0.9 $M_{\odot}$ for \BD. The couples of lines refer to two values of the wind velocity index $\beta$ = 0.6 and $\beta$ = 1. The horizontal dotted lines represent the minimum and maximum of the estimated luminosity of the blackbody component.}
\label{luminosita}
\end{figure}

As in the case of \HD, the ultraviolet spectra of \BD\ show N\,{\sc v} and C\,{\sc iv} resonance lines with P Cygni--like profiles, indicating the presence of a stellar wind. It is possible that part of the sdO stellar wind is captured by its compact companion, giving rise to accretion--powered X--ray emission. Since \BD\ does not fill its Roche lobe, we are in the case of wind accretion onto the compact object. Assuming that the Bondi--Hoyle scenario is applicable, we can estimate the expected luminosity. The density and velocity of the wind material at the position of the compact object can be computed assuming a canonical wind velocity law with radial dependence $v(R) = v_{\infty} (1-1/R)^{\beta}$, where $R$ is the radial distance in units of stellar radii and the index $\beta$ is typically in the range 0.6--1. In this way we can estimate the accretion luminosity as a function of the unknown orbital period, and compare it to the estimated range of the BB luminosity. In Fig.~\ref{luminosita} we report our estimates of the accretion luminosity in the cases of both a WD and a NS, together with the lower and upper limits for the BB luminosity. Although the observed luminosity is only poorly constrained, due to the large uncertainties in the spectral parameters, it is clearly consistent with that expected from a NS orbiting \BD\ with a period between about one and several days.

The blackbody emitting radius derived from the best fit is only marginally consistent with a neutron star; however, this parameter is strongly correlated with the poorly constrained slope of the power--law spectral component and an acceptable fit can be obtained, e.g., with $\Gamma$ = 1.5, $kT_{\rm BB}\sim$ 58 eV, and $R_{\rm BB}\sim$ 10 km. The observed spectrum is much softer than that typical of neutron stars in classical X--ray binaries, but these ones have very different companion stars and accrete at higher rates. Alternatively, the accreting companion could be a WD, if the sdO extends to (or close to) the Roche lobe, thus yielding an accretion rate larger than what we computed assuming wind accretion (or in the unlikely case that the adopted distance is largely overestimated).

\section{Summary and conclusions}

\begin{figure}[h]
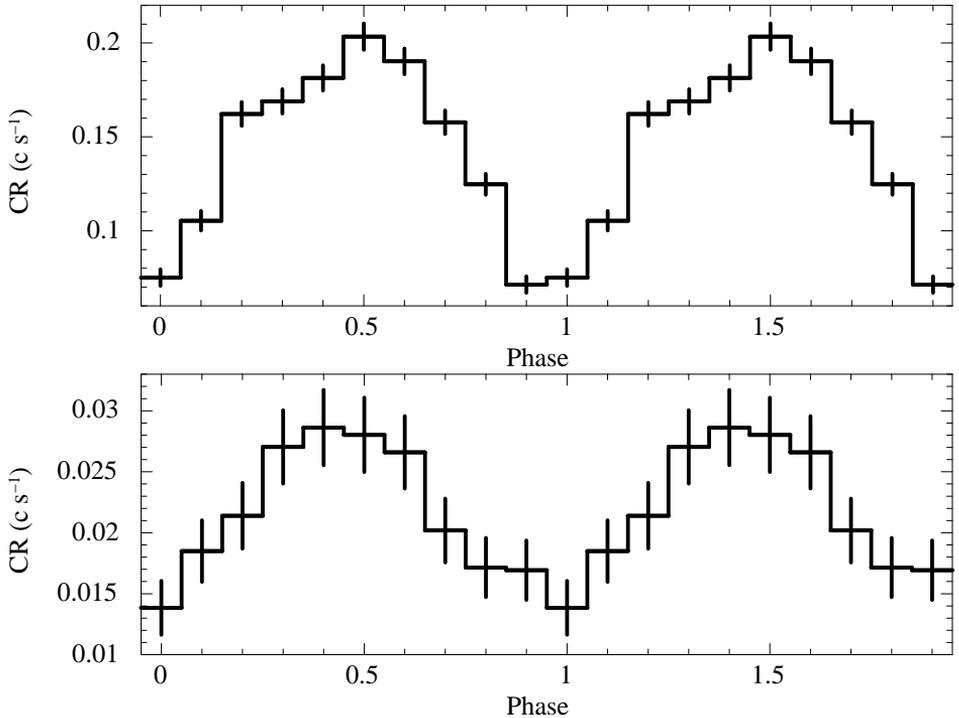

\centering
\resizebox{\hsize}{!}{\includegraphics[angle=-90,clip=true]{lc_HD.ps}}
\resizebox{\hsize}{!}{\includegraphics[angle=-90,clip=true]{lc_BD.ps}}
\caption{\XMM\ background--subtracted light curves of \HD\ (\textit{top}) and \BD\ (\textit{bottom}), in the energy range 0.15--2 keV, folded at the best--fit period \textit{P} = 13.18425 s and \textit{P} = 19.156 s, respectively.}
\label{timing}
\end{figure}

\begin{figure}[h]
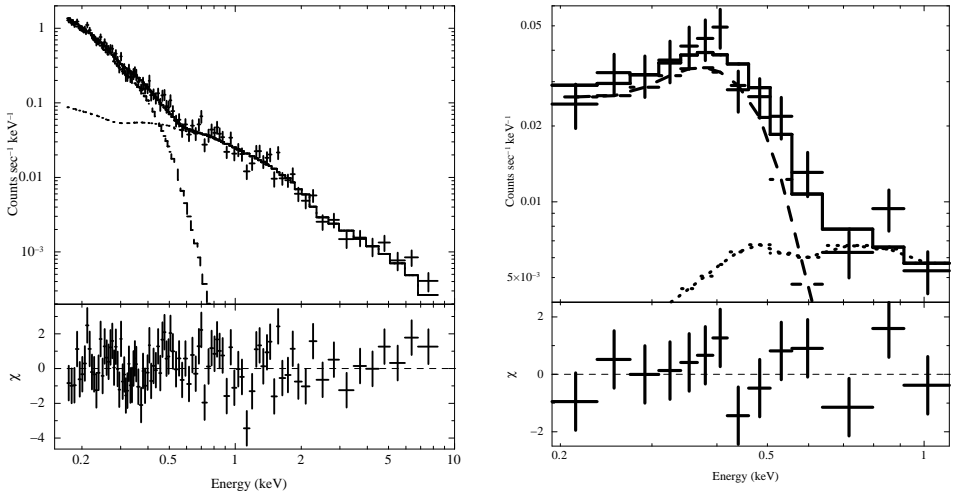

\centering
\begin{tabular}{c@{\hspace{1pc}}c}
\includegraphics[height=6.1cm,angle=-90]{spettro_HD.ps} &
\includegraphics[height=5.9cm,angle=-90]{Spettro_BD_new1.ps} \\
\end{tabular}
\caption{\textit{Top panels}: \XMM\ spectrum of \HD\ (\textit{left}) and \BD\ (\textit{right}) with the best--fit power--law (\textit{dotted line}) plus blackbody (\textit{dashed line}) model.
\textit{Bottom panel}: residuals (in units of $\sigma$) between data and
model.}
\label{spectral}
\end{figure}

The only two sdO stars observed with sensitive X--ray instruments have led to the discovery in both cases of a compact companion with very similar timing and spectral X-ray properties (Fig.~\ref{timing} and~\ref{spectral}). This is a noteworthy result, which might suggest that the fraction of sdO binaries is larger than currently believed on the basis of optical observations. We note that, contrary to late type companions, which can give a detectable contribution in the spectral and photometric data, WDs and NSs are too faint and completely outshined in the optical/UV by the sdO emission.

Both \BD\ and \HD\ belong to the subclass of luminous, He-rich sdOs, whose origin and evolutionary link with other classes of stars is still unclear \citep{nap08,jus11}. In this respect, X--ray observations might provide important information to complement optical/UV data, since the high sensitivity of current satellites like \XMM\ and Chandra can be used to find out compact companions of other sdO stars.

\section*{Acknowledgements} 
This work is based on observations obtained with \XMM, an ESA science mission with instruments and contributions directly funded by ESA Member States and NASA. We acknowledge financial contributions by the Italian Space Agency through ASI/INAF agreements I/009/10/0 (for the data analysis) and I/032/10/0 (for the \XMM\ operations).

\bibliographystyle{ceab}
\bibliography{Proceeding_LaPalombara_Subdwarfs_I2}

\end{document}